\documentclass[twocolumn,showpacs,preprintnumbers,amsmath,amssymb]{revtex4}
\usepackage{graphicx}

\begin{document}
\title{Multi-pronged events from Coulomb fission of nuclei at very low energies}
\author{B. Basu$^{1,2}$, S. Dey$^1$, D. Gupta$^{1,2}$, A.
Maulik$^1$, S. Raha$^{1,2}$, Swapan K. Saha$^{1,2}$, D. Syam$^3$}
\affiliation {$^1$Centre for Astroparticle Physics and Space
Science, Bose Institute, Kolkata 700091, India\\
$^2$Department of Physics, Bose Institute, Kolkata 700009, India\\
$^3$Department of Physics, Barasat Government College, Kolkata
700124, India}
\date{\today}

\begin{abstract}
Multi-pronged tracks have been recorded in the polyethylene
terephthalate (C$_{10}$H$_8$O$_4$)$_n$ solid state nuclear track
detector by exposure to a $^{252}$Cf fission source. After chemical
etching, two-prong to six-pronged tracks along with single tracks
have been observed under the optical microscope. We carried out a
systematic study to understand the origin of the prongs. The track
detectors were coated with metals (Cu, Ag and Pb) and were exposed
to $^{252}$Cf source. After chemical etching two-prong to four
pronged tracks were observed in each plate. We believe that at this
very low energy of the order of 1 MeV/A, Coulomb fission is the only
plausible explanation for the origin of such multi-pronged tracks.
\end{abstract}


\maketitle

The study of nuclear particles through track etching in solid state
nuclear track detectors (SSNTD) is widely practised~\cite{FL75,
DU87} in the field of cosmic rays. The polyethylene terephthalate
(C$_{10}$H$_8$O$_4$)$_n$ solid state nuclear track detector,
commercially known as PET, has been used for the detection of heavy
charged particles for many years \cite{BA05}. This detector has some
important advantages over the standard solid state nuclear track
detectors CR-39, Lexan etc. The PET has a much higher detection
threshold (Z/$\beta$ = 150) \cite{BA08} than that of CR-39
(Z/$\beta$ = 6) and Lexan (Z/$\beta$ = 57). The Z and $\beta$ are,
respectively, the charge and velocity ({\it v/c}) of a nuclear
particle incident on the track detector. Our earlier
studies~\cite{BA05} show that PET does not detect 6 MeV alpha
particles from $^{252}$Cf fission source. Further experiments
confirmed that it does not detect 650 keV (Z/$\beta$ = 107) alpha
particles which has the highest stopping power in the detector
material. The lower energy alpha particles were obtained by the
degradation of energy. It may be noted that efficient detection of
heavy cosmic ray particles requires that light charged particles
including protons and alpha particles be eliminated, which are
present as a large background. In this regard, the use of the PET
detector is very advantageous~\cite{BA082}. Another big advantage of
PET over CR-39 is that PET is a low cost detector and therefore very
useful when large area exposure is required. As a PET detector, we
have chosen a particular PET brand of CENTURY de'Smart, India, which
is commonly used as over-head projector transparencies (OHP).

\begin{figure}[h]
\includegraphics[width = 1.0\hsize,clip]{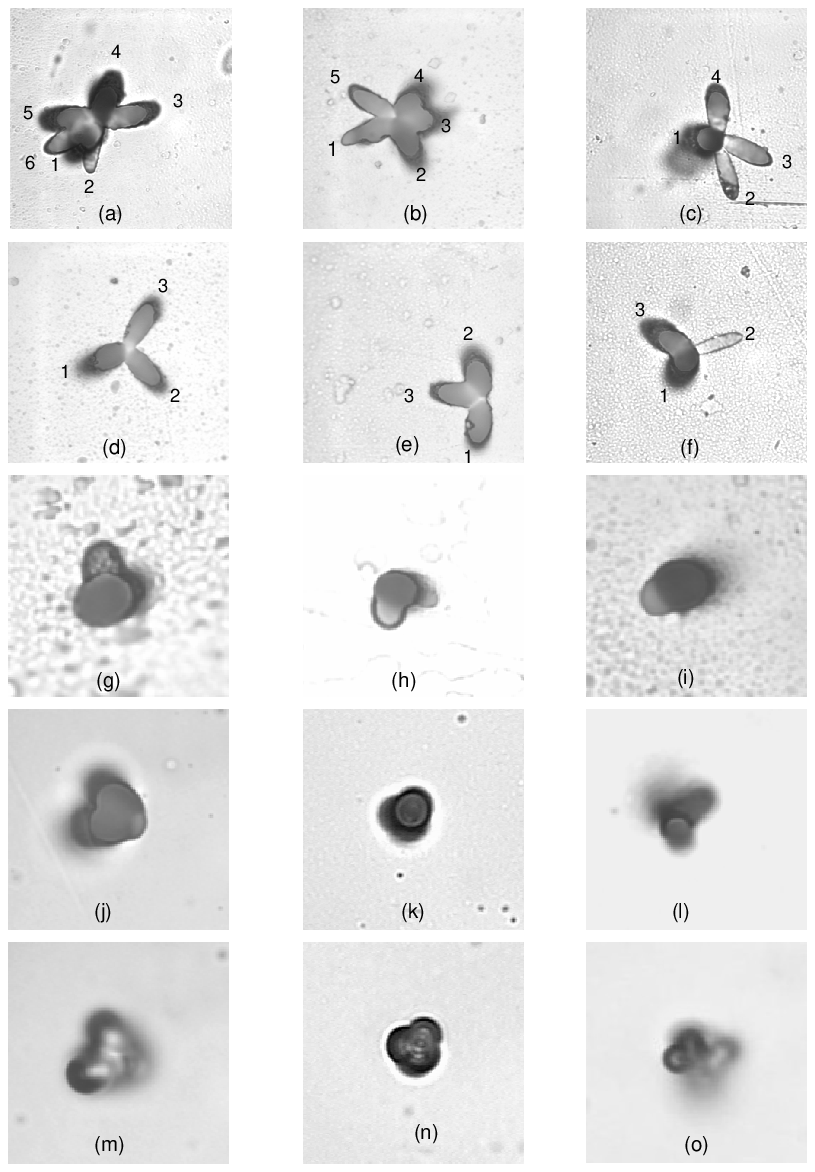}
\caption{Multi-pronged tracks in uncoated PET (a-f), Pb-coated PET
(g-i) and coated CR-39 on the surface (j-l) and at a depth of 3 $\mu
m$, 4 $\mu m$ and 4 $\mu m$ (m-o). The numbers in (a-f) correspond
to fragments, as depicted in Table 1.\\}
\end{figure}

During a systematic study of PET characteristics, it was exposed to
a $^{252}$Cf fission source. Two plates each of CR-39 and PET were
exposed to the fission source. One of the plates of PET detectors
registered clear multi-pronged charged particle tracks (Fig. 1)
along with single tracks, while the other detector registered only
single tracks of fission fragments. The CR-39 registered both
fission fragments and alpha particles and all are single tracks.
These multi-pronged tracks of nuclei have been studied with the help
of the calibration curve ($dE/dx$ vs $V{\rm _t}/V{\rm _g}$). The
$dE/dx$ is the stopping power of the charged particle in the
detector material, $V{\rm _t}$ is the track etch rate and $V{\rm_g}$
is the general etch rate. The $dE/dx$ values were obtained using
SRIM~\cite{ZI03}. The calibration curve was obtained from studies of
the tracks of different charged particles from radioactive sources
($^{252}$Cf) as well as ion beams from the accelerators. The charge
response of this PET detector to light nuclei has been studied using
3.4 MeV/A $^{16}$O-ions~\cite{BA08}, using the air gap between the
flange of the beam pipe and the detector as the energy degrader. The
charge response has also been studied for heavy nuclei using 11.1
MeV/A U-ion beam, with aluminium foils of different thickness to
degrade the incident energy to several values~\cite{BA08}.
Calibration has also been carried out using 3.9 MeV/A $^{32}$S and
2.7 MeV/A $^{56}$Fe beams. In the present work, the data reduction
was based on the measurement of the geometrical track parameters
(lengths, angles and etched pit diameters). The estimation of the
charge, mass and energy of the detected fragments show that the
fragments lie between Oxygen and Cobalt.

The CR-39 (Intercast Europe Co., Italy) and PET films, each of area
3.5 x 3.5 cm$^2$ were exposed in vacuum (10$^{-5}$ bar) for two
minutes to the incident alpha particles and fission fragments from a
$^{252}$Cf source of strength ~1 $\mu$Ci. The thickness of the PET
detector was 100 $\pm$ 5 $\mu$m and that of CR-39 was 600 $\mu$m.
The etching of exposed PET and CR-39 detectors were carried out in
6.25N NaOH solution at temperatures of 55$^0$C and 70$^0$C
respectively. Etching time was 4 hours which produced well developed
tracks. The plates were scanned under different dry objectives
(x100, x50, x20) of a Leica digital microscope, interfaced with a
computer for image analysis. Two to six pronged tracks were observed
in one of the PET films, along with single tracks (Fig. 1). The
multi-pronged tracks were observed in a localized area of the
detector. Figures of two-pronged tracks are not shown, as the tracks
may correspond to elastically scattered particles. Fig. 1 (a-f) show
spectacular multi-pronged events in uncoated PET. The numbers in
(a-f) correspond to fragments, as depicted in Table 1. Fragmentation
apparently resulted from the collision of the projectile with the
collimator (6 mm diameter, made of aluminium), shutter (made of
lead) or the target holder (made of aluminium). To confirm the
reactions we exposed a PET detector covered with 1 $\mu$m thick
aluminized mylar foil. After etching, it showed a few two-pronged
tracks, one three-pronged track and one four-pronged track.
Photographs of multi-pronged tracks show that the tracks originate
from single vertex. We have constructed the vertex for a few
three-pronged events. A least square-fit calculation yields a vertex
at a height of about 4 $\mu$m.  The thickness loss due to etching is
about 3 $\mu$m for four hours. Then it can be concluded that the
fragmentation occurred on the detector surface or just above it. The
observation of these multi-pronged tracks prompted us to initiate a
systematic study on PET and standardized CR-39 detectors by coating
them with different metals (coatings about 200 $\mu$g/cm$^2$) to see
if these fragments are the result of the presence of these metals.

Fig. 1 shows the projections of the tracks on a horizontal plane.
The actual angles between the directions of emission of the
fragments causing the tracks are given in Table 1, which also
includes fragment charge and energy. Fig. 1 (g-i) show tracks formed
in Pb-coated PET. We also used CR-39 coated with Cu, Ag and Pb. Fig.
1(j-l) show the tracks in Cu-, Ag-, Pb-coated CR-39 on the surface
of the detector as well as more prominent projections of the tracks
when focussed at 3 $\mu$m, 4 $\mu$m and 4 $\mu$m depth respectively
(Fig. 1 (m-o)).

To calculate the errors in the data, we find that the standard
deviation in the mean of the measurement of the depth of the tip of
the etch cone beneath the surface is about 10 $\%$ (depth resolution
of the microscope is 1 $\mu$m). The error in the general etch rate
$V{\rm_g}$, determined by the thickness loss method, is related to
the number of measurements made at various points of the PET sheet
and the average value of $V{\rm_g}$. Error is taken as the standard
deviation in the mean, which gives a value of about 5$\%$. Combining
these errors we get an error of about 10-15$\%$ in the determination
of $V{\rm _t}/V{\rm _g}$ which leads to similar errors in the
determination of $dE/dx$ from the $dE/dx$ vs $V{\rm _t}/V{\rm _g}$
calibration curve. The error in $dE/dx$ leads to an error of about
$\pm$ 3 in the determination of $Z$ from the curve between $dE/dx$
and $R$ obtained using SRIM~\cite{ZI03}, where $R$ is the range of
the fragment in the SSNTD. This error in $Z$ also leads to an error
in the determination of energies of the fragments up to a maximum of
about $\pm$ 30$\%$.

As to the origin of these multi-pronged events, it appears that at
this extremely low energy (energy of fission fragments being about 1
MeV/A) fragmentation of nuclei by the Coulomb force may be possible.
It was proposed~\cite{WI67,HO71,KR80} that the Coulomb field of a
heavy ion may distort a fissile nucleus so that it is slowly carried
over the barrier with little internal excitation. Since the
projectile energies are much lower than the Coulomb barrier, only
electric forces are involved in inducing the process. If we compare
Coulomb excitation with Coulomb fission, in the former case the
nucleus is shot up to an excited state. In the latter case the
nucleus is lifted through many states of collective excitation. If
it reaches the barrier state, fission occurs. The Coulomb field
actually exerts a torque which strongly aligns the nuclear symmetry
axis perpendicular to the projectile direction. The maximum
alignment occurs shortly after closest approach~\cite{KR80}.
Consequently, in the rest frame of the fragmenting nucleus, the
fission fragment distribution should peak at 90$^o$ relative to the
projectile direction for binary fission. For fission resulting in
three or more fragments the angles of emission can be quite
different from 90$^o$. Since in our case the projectiles are fission
fragments of $^{252}$Cf which are deformed and neutron rich, it is
the projectile rather than the target (plastic track detectors in
our case) which undergoes Coulomb fission. So in the lab frame the
peaking of the distribution should occur at angles less than 90$^o$.

We conclude by reiterating that the initial observation of the
multi-pronged tracks was in a plain (uncoated) PET detector exposed
to $^{252}$Cf source while no multi-pronged tracks were seen in
unexposed plates. In order to check this conclusion, we got the
plastic films coated with various metals before exposure to the
fission fragments. The metal-coated detectors indeed reproduce
multi-pronged tracks.

The present work may be the first observation of multi-pronged
events from Coulomb fission at such a low energy. Further studies in
this area as well as detailed theoretical analysis is presently
being pursued.

One of the authors (B.B) is thankful to the Department of Science
and Technology (DST), Government of India, New Delhi, for the award
of Women Scientist Project (No. SR/WOS-A/PS/-78/2003). The authors
thank Sujit K. Basu for technical assistance. This work has been
supported by the IRHPA (Intensification of Research in High Priority
Areas) Project (IR/S2/PF-01/2003) of the Science and Engineering
Council (SERC), DST, Government of India, New Delhi.

\begin{table}
\rotatebox{90}{%
Table 1: The charge, energy and angle between fragments. The first
and second columns correspond to pictures (a-f) for PET(uncoated),
as shown in Figure 1. }
\rotatebox{90}{%
\setlength{\tabcolsep}{3.5 mm}
\begin{tabular}{|c|c|c|c|c|c|c|c|c|c|c|c|c|c|l|}
  \hline\hline
  \multicolumn{1}{|c|}{Fig. 1}&
\multicolumn{1}{c|}{No. of fragments}& \multicolumn{12}{c|}{Charge
and energy (MeV) of the fragments}& \multicolumn{1}{c|}{Angle between tracks}\\
\multicolumn{1}{|c|}{}& \multicolumn{1}{c|}{}&
\multicolumn{2}{c|}{1}& \multicolumn{2}{c|}{2}& \multicolumn{2}{c|}{
3}&\multicolumn{2}{c|}{4}&\multicolumn{2}{c|}{5}&\multicolumn{2}{c|}{6}&\multicolumn{1}{c|}{}\\
   &  & Z & E & Z & E & Z & E & Z & E & Z & E & Z & E &\\
  \hline
  a & 6 & 10 & 18 & 14 & 35 & 14 & 37 & 14 & 40 & 16 & 45 & 11 & 20 & $\theta_{12}$=31$^o$, $\theta_{23}$=105$^o$, $\theta_{34}$=54$^o$\\
  &&&&&&&&&&&&&&$\theta_{45}$=83$^o$, $\theta_{56}$=31$^o$, $\theta_{61}$=25$^o$\\
\hline
  b & 5 & 18 & 60 & 20 & 30 & 17 & 50 & 9 & 16 & 13 & 30 &  &  & $\theta_{12}$=86$^o$, $\theta_{23}$=52$^o$, $\theta_{34}$=37$^o$ \\
  &&&&&&&&&&&&&&$\theta_{45}$=83$^o$, $\theta_{51}$=54$^o$\\
\hline
  c & 4 & 21 & 65 & 12 & 25 & 16 & 45 & 17 & 50 &  &  &  &  & $\theta_{12}$=87$^o$, $\theta_{23}$=50$^o$, $\theta_{34}$=100$^o$ \\
  &&&&&&&&&&&&&&$\theta_{41}$=62$^o$\\
\hline
  d & 3 & 13 & 25 & 15 & 43 & 19 & 52 &  &  &  &  &  &  & $\theta_{12}$=99$^o$, $\theta_{23}$=93$^o$, $\theta_{31}$=83$^o$ \\
\hline
  e & 3 & 12 & 20 & 10 & 15 & 10 & 15 &  &  &  &  &  &  & $\theta_{12}$=108$^o$, $\theta_{23}$=58$^o$, $\theta_{31}$=73$^o$  \\
\hline
  f & 3 & 12 & 20 & 10 & 17 & 11 & 18 &  &  &  &  &  &  & $\theta_{12}$=101$^o$, $\theta_{23}$=105$^o$, $\theta_{31}$=58$^o$ \\
  \hline\hline
\end{tabular}
}
\end{table}

\end{document}